\documentclass[10pt,letterpaper,twocolumn]{article} 

\usepackage{ol2}
\usepackage[draft]{hyperref}
\usepackage{amsmath}
\usepackage{bm,braket}

\begin{document}

\abovedisplayskip=6pt
\abovedisplayshortskip=6pt
\belowdisplayskip=6pt
\belowdisplayshortskip=6pt
\twocolumn[ 

\title{Design of an efficient single photon source from a metallic nanorod dimer: a quasinormal mode finite-difference time-domain approach}


\author{Rong-Chun Ge$^{*}$  and S. Hughes}
\address{
Department of Physics, Engineering Physics and Astronomy,
Queen's University, Kingston, Ontario, Canada K7L 3N6
$^*$Corresponding author: rchge@physics.queensu.ca
}

\begin{abstract}
We  describe how the finite-difference time-domain (FDTD) technique can be used to compute the   quasinormal mode (QNM) for metallic nano-resonators, which is important for describing and understanding light-matter interactions in nanoplasmonics. 
We  use the QNM to model the enhanced  spontaneous emission rate for dipole emitters near a  gold nanorod  dimer structure  using a newly developed QNM expansion technique. Significant enhanced photon emission factors of around 1500 are obtained with large output $\beta$-factors of about $60\%$. 
\end{abstract}

 ] 

\noindent
Resonant cavity structures have the ability to trap light in very small  spatial volumes
which has a wide range of applications in nanophotonics \cite{ChangBook}.
Various miniaturized 
 cavity structures have been developed over the years to  manipulate  light at the subwavelength scale, and extreme nanoscale confinement is  now possible with  metallic nano-resonators (MNRs). In a frequency regime  near  a localized surface plasmon (LSP) \cite{plasmonBook}, the local density of optical states (LDOS)  can be increased dramatically. Consequently, the spontaneous emission (SE) rate  of a dipole emitter can be enhanced via the Purcell effect. The resonant enhancement from MNRs has  applications in  
  chemical sensing~\cite{KR07}, high resolution imaging~\cite{KYHLIRM97,Hou:13},
 optical antennas~\cite{PBL09} and   single photon emission~\cite{Maitre13}.

While the optical properties of MNRs are being actively pursued,
numerically modelling of the basic cavity physics is  extremely demanding,
and analytical solutions of the  modes only exists for very simple
structures such as  spheres. 
For resonant cavity structures,  the natural modes of the system
are  called quasinormal modes (QNMs) \cite{Lee99, Leung941},
defined as the frequency domain solutions to the wave equation with open boundary conditions (the Silver-M\"uller radiation condition).
Kristensen {\em et al.}~\cite{Philip:OL12} first  used the QNMs  to introduce a rigorous definition of the ``generalized effective mode volume'' and Purcell factor \cite{Purcell}, and applied these results to photonic cavity structures. For MNRs, the QNMs 
also form the natural starting point for developing  analytical theories of light-matter interactions in nanoplasmonics \cite{Sauvan13,ACSPerspectives,USPRL}.

One of the most common numerical techniques for obtaining the cavity mode
for dielectric cavities  is the finite-difference time-domain (FDTD) technique~\cite{cFDTD}. The FDTD technique allows one to simulate open boundary conditions with ``perfectly matched layers'' (PMLs) located at the  leaky mode region outside the cavity.
For dielectric cavities,
this open-boundary FDTD approach has been shown to yield excellent agreement with direct integral 
equation methods \cite{Philip:OL12}. Other time-domain techniques such as the discontinuous Galerkin time-domain approach can also use PMLs \cite{Busch1}.
For metals, a major problem occurs when using the usual mode calculation approach with a dipole excitation source, since the extracted mode depends sensitively on the dipole position \cite{HuangOE2011}, and is therefore incorrect. Thus it has been common practice to excite the MNR with a plane wave
source and obtain the {\em scattered field}. However, this scattered field is not the same field as the QNM and it cannot be properly normalized for use
in quantum optics, e.g., for obtaining the Purcell factor and effective mode volume  \cite{ACSPerspectives}---two well known quantities that help describe the underlying physics of cavity light-matter interactions.
While some frequency-domain techniques exist for computing the 
QNMs of MNRs~\cite{BPSHL13,LMK13}, it is highly desirable to be able to compute the QNMs using the commonly employed and general FDTD technique.

The FDTD method is already widely used by the plasmonics community, and its accuracy for obtaining the enhanced field has been verified
against other numerical techniques such as the multipole expansion technique ~\cite{ASMT09}.
 In addition, the LDOS can be  calculated by employing a dipole excitation source~\cite{VS07,YaoLaserReviews,ColeOL},
which can also model {\em local field effects}, e.g.,  associated with finite-size photon emitters
inside a MNR \cite{ColeOL}. While direct dipole calculations are feasible, they are very time consuming and require a new simulation (which may require many hours of computational time) for each spatial position of the dipole emitter; thus a QNM picture would be much more efficient, since it allows one to simulate dipole responses both as a function of space and frequency as soon as the QNM is obtained and properly normalized.  

In this Letter  we first describe how the FDTD technique \cite{FDTDs} can be efficiently employed to obtain  QNMs of a MNR by filtering the scattered field with a temporal window function.
 We compute the
spatial dependence of the QNM and the effective mode volume,
and  show excellent agreement with full-dipole calculations
for the enhanced SE factor of a dipole emitter.  We then show how
a  gold nanorod dimer can act as an efficient single photon source with large Purcell factors (1500) and impressive output  $\beta$-factors
(around 60\%). {In contrast to spherical dimer structures~\cite{aplsun}, we find that the nanorod dimer acts to increase the  $\beta$-factor
for good photon emitters (in comparison to a single resonator), and has the additional advantage
of yielding resonant frequencies in the visible spectrum.}

The QNM $\tilde{\bf f}_\mu$ has a complex eigenfrequency, 
 $\tilde{\omega}_{\mu}=\omega_\mu-{\rm i}\gamma_\mu$, with a quality factor $Q=\omega_\mu/2\gamma_\mu$. The QNM is normalized through~\cite{Leung941,Lee99}
\begin{align}
\langle\langle \tilde{\bf f}_{\mu}|\tilde{\bf f}_{\nu}\rangle\rangle\!&=\!\lim_{V\rightarrow\infty}\int_V\left(\frac{1}{2\omega}\frac{\partial (\epsilon({\bf r},\omega)\omega^2)}{\partial \omega}\right)_{\omega=\tilde{\omega}_{\mu}}\!\!\!\!\tilde{\bf f}_{\mu}({\bf r})\cdot\tilde{\bf f}_{\nu}({\bf r})d{\bf r} \nonumber\\
&+ \frac{ic}{2\tilde{\omega}_{\mu}}\int_{\partial V}\sqrt{\epsilon({\bf r})}\tilde{\bf f}_{\mu}({\bf r})\cdot\tilde{\bf f}_{\nu}({\bf r})d{\bf r}=\delta_{\mu\nu}.
\label{eq:norm}
\end{align}
Since the eigenfrequencies are complex, the QNMs diverge in space and each part of Eq.~(\ref{eq:norm}) diverges,
but the  total sum converges quickly in space \cite{ACSPerspectives,USPRL}.
This convergence occurs approximately when the outgoing field becomes purely oscillatory, rather than evanescent.
For spatial points near the resonator,
the transverse part of the photon Green function can be expanded as~\cite{Lee99,USPRL}
${\bf G}^{\rm T}({\bf r}_1,{\bf r}_2;\omega)= \sum_{\mu}\frac{\omega^2}{2\tilde\omega_{\mu}(\tilde\omega_{\mu}-\omega)}\tilde{\bf f}_{\mu}({\bf r}_1)\tilde{\bf f}_{\mu}({\bf r}_2)$.
One can then derive the effective mode volume
$V_{\rm eff}$ and  enhanced SE factor
$F_a({\bf r}_{ a},\omega)$, where ${\bf r}_{ a}$ is the spatial position of a dipole emitter. 
For dipole positions near the resonator,
these quantities are defined through~\cite{Philip:OL12,USPRL} 
\begin{align}
\frac{1}{V_{\rm eff}}  = \text{Re}\left\{\frac{1}{v_\text{Q}} \right\},\quad v_\text{Q} = \frac{\langle\langle\tilde{\bf f}_\text{c}|\tilde{\bf f}_\text{c}\rangle\rangle}{\varepsilon_{\rm B}\tilde{\bf f}_\text{c}^2({\bf r}_0)},
\label{eq:Vq}
\end{align}
and 
\begin{align}
F_a({\bf r}_{ a},\omega) =
F_\text{P}\,\eta({\bf r}_{ a},{\bf n}_{ a};\omega) +1,
\label{eq:Fq}
\end{align}
where
$
F_\text{P}\,\eta({\bf r}_{\rm a})=
\frac{3Q}{4\pi^2}\frac{\lambda_\text{c}^3}{n_\text{B}^3}\frac{\omega_\text{c}^2\gamma_\text{c}}{\omega}
\text{Im}\left[\!
\frac{\mathbf{n}_{ a}\!\cdot\!\varepsilon_\text{B}\tilde{\bf f}_{\rm c}({\bf r}_{ a})\tilde{\bf f}_{\rm c}({\bf r}_{ a})\!\cdot\!\mathbf{n}_{ a}}{\tilde{\omega}_{\rm c}(\tilde{\omega}_{\rm c}-\omega)}
\!\right ]
$
is the Purcell factor, $F_\text{P}$, multiplied by a factor $\eta$ to account for any deviations at ${\bf r}_{ a}$ from the field maximum ${\bf r}_0$, cavity polarization, and cavity resonant frequency~\cite{USPRL}; here 
 ${\bf n}_a$ is the unit vector in the direction of the dipole emitter,
$n_\text{B}$ is the background-medium refractive index  in which the MNR is embedded and $\lambda_c$ is the corresponding  wavelength.
Equation (\ref{eq:Fq}) assumes a single QNM expansion, which 
is valid for the MNR light-matter interaction regimes that we study below.

The FDTD technique can also obtain the enhanced SE factor directly, but {\em only} at the dipole position, which can be used to check the numerical accuracy of the  QNM equations. Specifically, one can  define \cite{NovotnyAndHecht_2006}
\begin{align}
F^{\rm FDTD}_{a}({\bf r}_a,\omega) 
= \frac{\text{Im}\left\{{\bf n}_a\cdot{\bf G}^{\rm FDTD}({\bf r}_{a},{\bf r}_a;\omega)\cdot{\bf n}_a\right\}}{\text{Im}
\left\{{\bf n}_a\cdot{\bf G}_\text{B}
({\bf r}_a,{\bf r}_a;\omega)\cdot
{\bf n}_a\right\}},
\label{Eq:PurcellFromLDOS}
\end{align}
where ${\bf G}_{\rm B}$  is the  known  homogeneous medium  Green function 
\cite{NovotnyAndHecht_2006}, and ${\bf G}^{\rm FDTD}({\bf r}_{a},{\bf r}_a;\omega)$
is obtained from FDTD  at the dipole source position \cite{YaoLaserReviews}. Note  that our full-dipole FDTD calculations have been shown elsewhere to yield
 good agreement with analytical results for metal spheres  \cite{ColePRB12}  (within $5\%$ agreement over broadband frequencies and various spatial positions).

Next we  discuss how FDTD is usually used to obtain resonator cavity modes and highlight the main problem for metallic structures.
As a direct space-time simulation method, FDTD can obtain the spectral response of system from a Fourier transform of its time response. Thus for any incident field with a finite bandwidth, the computed scattered field, ${\bf E}^{\rm S}(\omega)$, will contain the resonant responses of the system, e.g., spectral peaks corresponding to the QMNs. Defining the total electric field as
 ${\bf E}^{\rm tot}(\omega)$ (including the incident source),
 then the scattered field is simply 
  ${\bf E}^{\rm S}(\omega) = {\bf E}^{\rm tot}(\omega) - {\bf E}^{\rm h}(\omega)$, where ${\bf E}^{\rm h}(\omega)$ is the homogeneous solution without the MNR (i.e., no scattering).
 For a typical dielectric cavity with a high $Q$ resonance, almost any incident field or dipole source field can be used to obtain the mode of interested if one applies a temporal window to subtract off the excitation source. A common method of choice is to use a dipole excitation source and begin the Fourier transform of the scattered field after the excitation pulse has gone, which has been shown to yield  good agreement with other mode solving techniques \cite{Philip:OL12}. However, for a metal, the dipole source does not efficiently excite the QNM because of large losses,
and  extra care is needed in applying a temporal window function to compute the QNM of interest. Below we describe how the FDTD technique can obtain the true QNM, if the QNM is spectrally well isolated.

To cover a broad spectral range, we use a plane-wave excitation source with a 6~fs  time duration. To efficiently excite the QNM, we choose the polarization of the source field to be parallel to
the expected polarization of the QNM (e.g., along the rod axis for a nanorod). After obtaining the scattered field, we compute the QNM from\begin{align}
\tilde{\bf f}_{\rm c}({\bf r};\omega_{\mu})=\int_{0}^{t_{\rm end}} {\bf E}^{\rm S}({\bf r},t)e^{i\omega_{\rm c} t}e^{-\frac{(t-t_{\rm off})^2}{2(\tau_{\rm win})^2}}dt,
\label{eq:q1}
\end{align}
where $\omega_{\rm c}$ is the real part of the eigenfrequency $\tilde{\omega}_{\rm c}$, $2\sqrt{\ln2}\,\tau_{\rm win}$ gives the FWHM (full-width at half-maximum) of the time window, and $t_{\rm off}$ is the time offset from the center of the  source pulse 
 to the center of the time windowing function.
One criterion for the selection of time window parameters is to set $\omega_{\rm c} \tau_{\rm win}^2/2t_{\rm off}\rightarrow Q$, while keeping $\tau_{\rm win}$ as large as possible. To understand why this works, consider the ideal case for which the integral in Eq.~(\ref{eq:q1}) is carried out from $-\infty$ to $\infty$, and assume the scattered field is given by ${\bf E}^{\rm S}(t) = \sum_i{\bf E}_ie^{(-i\omega_i-\gamma_i)t}$ ($(\omega_i,\gamma_i)\neq(\omega_j,\gamma_j)$ if $i\neq j$);  after performing the integration, the amplitude of the $i$th component is proportional to $|{\bf  E}_i e^{\frac{({\rm i}\Delta\omega_i-\Delta\gamma_i)^2\tau_{\rm win}^2}{2}}|e^{-(\frac{t_{\rm off}}{\tau_{\rm win}})^2}$ with $\Delta\omega_i = \omega_i-\omega_{\rm c}$ and $\Delta\gamma_i = \gamma_i-t_{\rm off}/\tau_{\rm win}^2$. If  the resonances of the system are well separated, so that $|\Delta\omega_i|\tau_{\rm win}\gg 1,$ and  $|\Delta\omega_i|\gg|\Delta\gamma_i|$  for $i\neq{\rm c}$, then the QNM resonance, $\tilde{\omega}_{\rm c} = \omega_{\rm c} - {\rm i}\gamma_{\rm c}$,  will be the only surviving term after applying the windowing function. This is   achieved by having $t_{\rm off}/\tau_{\rm win}^2 = \gamma_\mu$ for the QNM of interest; however, due to the factor $e^{-\frac{t_{\rm off}}{\tau_{\rm win}}}$, $t_{\rm off}$ should not be too large, otherwise the spectrum will be too weak and influenced by  numerical errors. We have found that a good choice is to simply choose $t_{\rm off} = 4\pi/\gamma_{\rm c}$, i.e., two times the lifespan of the QNM.

\begin{figure}[t!]
\centering\includegraphics[trim=1.4cm 0.1cm 3.9cm 1.1cm, clip=true, width=0.9\columnwidth]{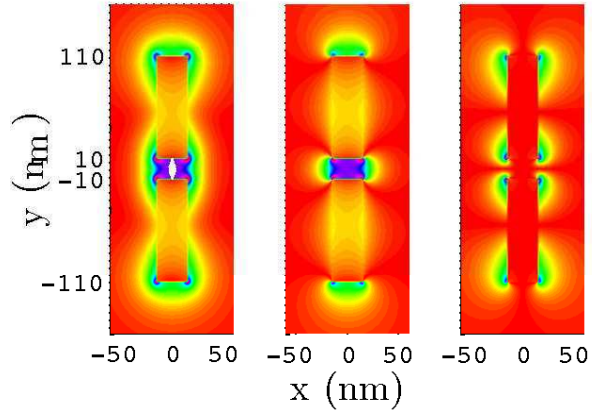}
\caption{Close up view of  the QNM profile of the gold dimer. Left: $|\tilde{\bf f}({ x,y,0};\omega_{\rm c})|$, middle: $|\tilde{\bf f}_y({ x,y,0};\omega_{\rm c})|, $ and right:  $|\tilde{\bf f}_x({ x,y,0};\omega_{\rm c})|$ at $\omega_{\rm c} = 291.06~$THz. A $y$-polarized  dipole  at the center of the gap is shown by the white dot and double arrow in the left panel.}
\label{f:f1b}
\end{figure}

Next we  apply the above technique to investigate  a  gold   dimer structure made up of two identical nanorods.
We choose a rod radius ${\rm r}_a = 15~$nm with an axis length $l = 100~$nm (along  $y$).
  We use the Drude model, $\varepsilon(\omega) = 1-\frac{\omega_{\rm p}^2}{\omega(\omega+{\rm i}\gamma)}$, with parameters similar to gold, with $\omega_{\rm p} = 1.26\times 10^{16}~$rad/s (plasmon frequency), and $\gamma = 1.41\times 10^{14}~$rad/s (collision rate).
 In order to get a  larger enhancement of the SE  for a quantum dipole, we  consider two nanorods that are parallel with each other.  The separation gap is set to  20~nm, which helps minimizes nonradiative decay and also allows sufficient space to embed a quantum emitter such as a quantum dot  (see
dipole arrow in Fig.~\ref{f:f1b}). From FDTD analysis, we  find the dipole mode is around $\tilde{\omega}_{\rm c}/2\pi=291.06 -{\rm i}20.28 ~$THz ($\omega_{\rm c}\approx 1.2~$eV), which is redshifted (by about 34~THz) with respect to a single nanorod \cite{USPRL} due to the bonding effect; the corresponding quality factor is $Q \approx 7.2$, which is smaller than that for  a   single nanorod ($Q_{\rm single}\approx 10$).

\begin{figure}[t!]
\centering\includegraphics[trim=0.2cm 0.3cm 0.3cm 1.2cm, clip=true, width=0.9\columnwidth]{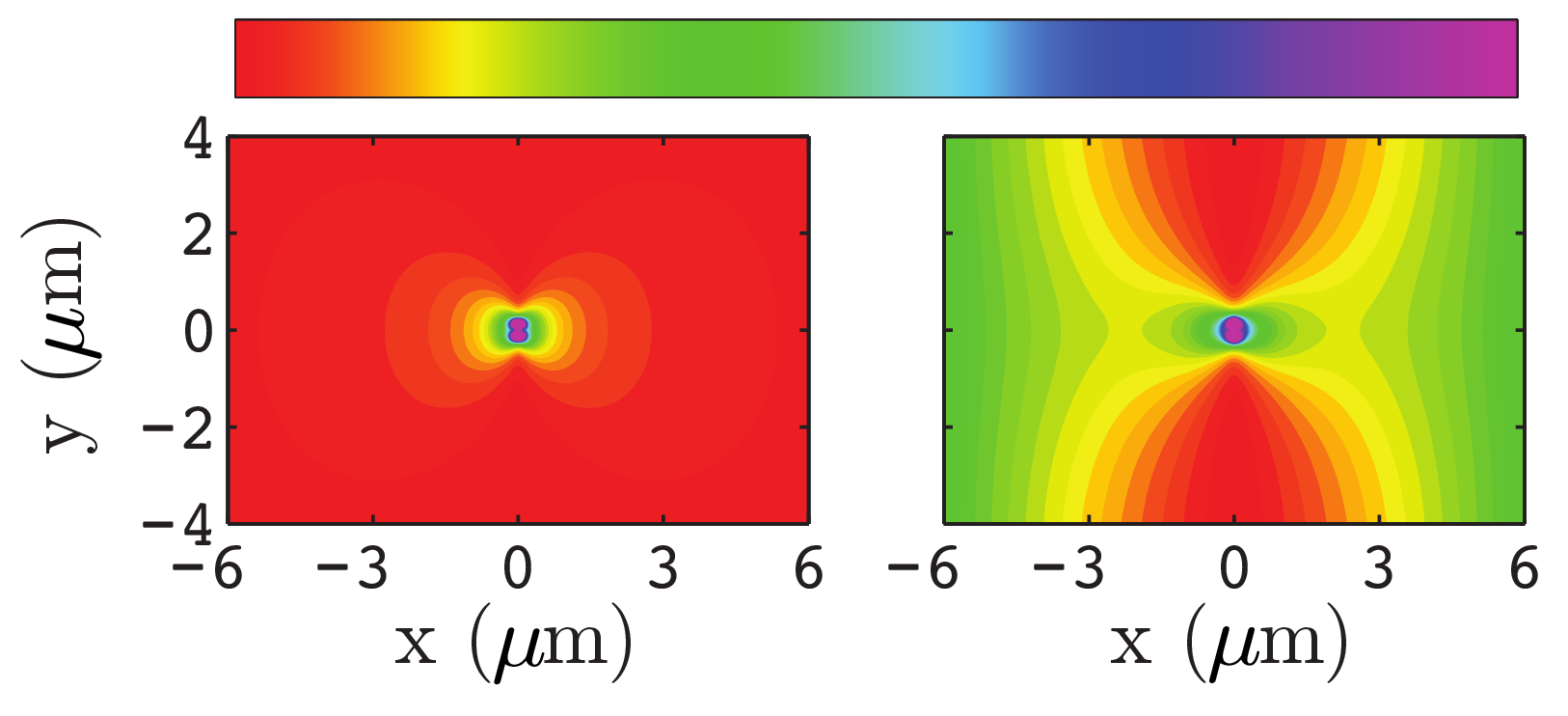}
\caption{ Spatial profile of the dipole mode of a gold nanorod dimer, showing both the scattered field and  QNM.    Left:
 $|{\bf E}^{\rm S}({ x,y,0};\omega_{\rm c})|$, and right: $|\tilde{\bf f}({ x,y,0};\omega_{\rm c})|$ at $\omega_{\rm c} = 291.06~$THz.
The excitation source is a $y$-polarized plane wave incident in the $x$ direction.}
\label{f:f1}
\end{figure}

To obtain the scattered field and the QNM, a $y$-polarized plane wave with frequency $\omega/2\pi = 291.06~$THz is employed, incident in the
$x$ direction. The simulation domain size is $12\times8\times6~{\rm \mu m^3}$ (Fig.~\ref{f:f1}) and $2.4\times2.4\times2.4~{\rm \mu m^3}$  (Fig.~\ref{f:f1b}), and we use a conformal meshing scheme with a maximum step size 40~nm in all  directions; a smaller refined mesh of 1~nm is used around the metal dimer;
%
100 layers of PML have been used with symmetric (antisymmetric) boundary condition in $z$ ($y$) direction for both simulation and the time step is 1.8875 as. 
We  use a time window with parameters $t_{\rm off} = 100~$fs and $\tau_{\rm win} = 23.3/\sqrt{\ln2}~$fs.
The left and right panels of Fig.~\ref{f:f1} show the scattered field and
QNM, respectively. As anticipated, the QNM is seen to have an increasing field value for positions
further away from the resonator, which is caused by the outgoing boundary conditions and the complex eigenfrequency.
A close-up view of the QNM near the resonator is shown in  the left panel of Fig.~\ref{f:f1b}, and the correspondent mode volume is calculated to be $V_{\rm eff} \approx 1.9\times 10^{-4}(\lambda_{\rm c}/n_B)^3$ (which is about double that of the single nanorod).  The oscillating charge distribution is of opposite sign at both ends of each nanorod; also, the oscillating charge at the bottom of the upper nanorod has a different sign from the charge at the top of the lower nanorod, which makes the localized surface plasmons of the individual nanorods  couple effectively. Consequently, the field inside the cavity (gap) is significantly enhanced and $y$-polarized.  The middle and right panels in  Fig.~\ref{f:f1b} show the $y$ and $x$ components of the QNM at $z=0$.
The node of the $y$-component sits around both ends of the nanorods, slightly inside this metal.
Similar  nodal lines  appear for an electric dipole composed of charge $\mp q$.

We remark that for these calculations it is important to choose
a $y$-polarized plane wave, which efficiently excites the QNM of interest.
If we use an $x$-polarized plane wave, with  the parameters same as above,
then a rather strange pseudo-mode  is obtained. 
The failure of using $x$-polarized plane wave to obtain the QNM can be explained by the fact that the QNM can not be efficiently excited since the dipole moment lies along the $y$ direction. So some care is needed in choosing the correct excitation source, though in practice this is easy to do when exciting the QNM of interest. We have also checked that our mode calculation approach
successfully  reproduces the mode profile and QNMs for dielectric cavity structures, e.g., in Ref.~\onlinecite{Philip:OL12} and also for other MNRs.

\begin{figure}[t!]
\centering\includegraphics[width=.85\columnwidth]{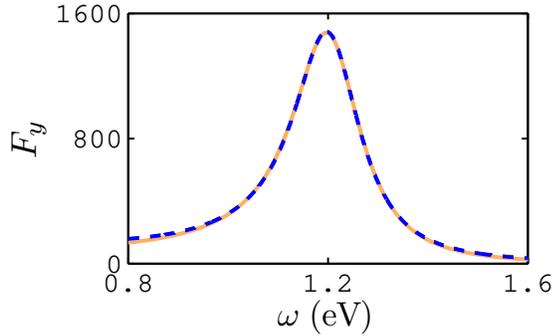}
\caption{Enhanced SE factor for  the gold nanorod dimer structure, with the blue (dashed) curve given by direct FDTD calculation (Eq.~(\ref{Eq:PurcellFromLDOS})) and the orange (solid) curve given by the mode expansion technique (Eq.~(\ref{eq:Fq})). The emission dipole is $y$-polarized  at the center of the gap of the dimer (see left panel of Fig.~\ref{f:f1b}.) }
\label{f3D1}
\end{figure}


{\it Single photon enhancement factors}.
As a possible application of using MNRs for single photon emitters, a large enhancement of the SE factor is desired. This can be achieved by  having a quantum dipole at the gap center of the dimer.
In Fig.~\ref{f3D1}, the enhanced SE factor, $F_y$, 
of a $y$-polarized  dipole emitter at the center of the gap is shown by the blue (dashed) curve using a full-dipole numerical calculation with no approximations, i.e., $F_{y}^{\rm FDTD}$~\cite{VS07,YaoLaserReviews}.
Next we use the QNM to obtain  the enhanced SE factor \cite{USPRL}. The result is shown by the orange (solid) curve in Fig.~\ref{f3D1};  it agrees extremely well with the full-dipole FDTD calculations. 
Indeed, the agreement also confirms that
the entire response is dominated by a single QNM which is highly desired for single photon source applications. 
From our results, we obtain impressive peak emission factors of around 1500.
Note that although the $Q/V$ factor of the dimer structure
is about three times {\em smaller} than the single nanorod, the emission factor of the dimer is 
around three times larger than for a single nanorod at an
equivalent dipole position. Thus the
$Q/V$ is clearly not the main metric
to describe the enhanced emission factors for dipole
away from the field antinode; note the reason for not choosing the field
antinode, is that the emission would be severely quenched from Ohmic heating and a single mode Purcell factor ceases to have any meaning  \cite{USPRL}.
In this regard, the other important figure-of-merit for emitting single photons is the output-coupling  $\beta$ factor (probability of photon emission via radiative decay {in the far field}) of the dipole emitter, which is computed here to be  $\beta \approx 58\%$; this again is better than the single gold nanorod, by about 10\%; {with respect to single nanorod, the electric field of the dimer is repelled outside the lossy nanorods (especially around the hot spots) due to opposite charge distribution between them. This leads an  overall nonradiative decay reduction ($\propto\int{\rm Im}[\epsilon({\bf r},\omega)]|{\bf E^{\rm tot}({\bf r},\omega)}|^2dV$), even though the volume of the integral is doubled.}

In summary, we have described how the commonly employed FDTD method can be used to efficiently  obtain the QNMs for
 metallic resonators by applying a filtering function to the scattered field.
These calculations are verified by comparing the 
resulting SE emission factor with
full-dipole FDTD numerical calculations, where we find
excellent  agreement. Using this technique, we
have proposed a metal nanorod dimer structure that acts  an efficient single photon source, yielding  a
 large Purcell factor  of 1500 and an  output $\beta$ factor of around 60\%.

This work was supported by the Natural Sciences and
Engineering Research Council of Canada. We thank Jeff Young and
Philip Kristensen for useful discussions.


\end{document}